# Optical spin pumping induced pseudo-magnetic field in two dimensional heterostructures

Chongyun Jiang[1†], Abdullah Rasmita[1†], Weigao Xu[1], Atac Imamoğlu[2], Qihua Xiong[1,3,4*], Wei-bo Gao[1,4,5*]

[1]Division of Physics and Applied Physics, School of Physical and Mathematical Sciences, Nanyang Technological University, Singapore 637371, Singapore.

[2]Institute of Quantum Electronics, ETH Zürich, CH-8093 Zürich, Switzerland

[3]NOVITAS, Nanoelectronics Center of Excellence, School of Electrical and Electronic Engineering, Nanyang Technological University, Singapore 637371, Singapore.

[4]MajuLab, CNRS-Université de Nice-NUS-NTU International Joint Research Unit UMI 3654, Singapore.

[5]The Photonics Institute and Centre for Disruptive Photonic Technologies, Nanyang Technological University, Singapore 637371, Singapore.

*Corresponding author. Email: qihua@ntu.edu.sg; wbgao@ntu.edu.sg.

†These authors contributed equally to this work

**Two dimensional heterostructures are likely to provide new avenues for the manipulation of magnetization that is crucial for spintronics or magnetoelectronics. Here, we demonstrate that optical spin pumping can generate a large effective magnetic field in two dimensional $MoSe_2$/$WSe_2$ heterostructures. We determine the strength of the generated field by polarization-resolved measurement of the interlayer exciton photoluminescence spectrum: the measured splitting exceeding 10 milli-electron volts (meV) between the emission originating from the two valleys corresponds to an effective magnetic field of ~ 30 T. The strength of this optically induced field can be controlled by the excitation light polarization. Our finding opens up new possibilities for optically controlled spintronic devices based on van der Waals heterostructures.**

Magnetic fields or pseudo-magnetic field interacting with low-dimensional system not only shows rich phenomena in condensed matter physics [1,2], but are also likely to play a key role in realizing the technological promise of spin-based devices [3-5]. In direct analogy with spintronics, the coupling of valley degree of freedom with magnetic field in atomically thin layer of graphene [6] or transition metal dichalcogenides (TMDs) provides new perspective towards the emerging field of valleytronics [7], as shown in cases of valley-polarized Landau levels [8,9], magnetic proximity effect [10,11], valley Zeeman splitting [12,13] and giant spin/valley susceptibility of TMD monolayers[14].

The introduction of magnetic field in 2D materials can be done in different ways either internally or externally. Internally, intrinsic magnetism has been demonstrated in two dimensional crystals $CrI_3$ [15] or $Cr_2Ge_2Te_6$ atomic layers [16]. Externally, the pseudo-magnetic field can be generated by other stimuli such as strain [17], molecular doping [18], and light [19-24]. Regarding previous pioneer works using light, all of the reported approaches require ultrafast lasers with high peak power for realizing inverse Faraday effect [19-21], AC stark shift [22,23] or recently Bloch-Siegert shift [24].

Here, we demonstrate a new way to generate an optically induced pseudo-magnetic field by optical spin pumping mechanism in 2D TMD heterostructures. The introduction of magnetic field relies on the ultrafast charge transfer [25, 26] and different spin polarization rate in different layers of the heterostructures. The coupling between this field and the interlayer exciton results in optically detectable valley degeneracy breaking. We observe a spectrum splitting between different valley transitions of more than 10 meV which corresponds to ~30 T pseudo-magnetic field. We shows that the degeneracy between the valleys can be lifted by using continuous wave optical excitation and that the valley splitting magnitude can be controlled by changing the excitation laser degree of circular polarization.

Our sample consists of a $MoSe_2$/$WSe_2$ heterostructure on $SiO_2$/Si substrate with the $MoSe_2$ monolayer stacked on top of the $WSe_2$ monolayer [27, 28]. Due to the band alignment of the $MoSe_2$/$WSe_2$ heterostructure, electrons and holes excited with light will relax into the conduction band of $MoSe_2$ and the valence band of $WSe_2$ respectively after a short time. Electrons in $MoSe_2$ and holes in $WSe_2$ bound by Coulomb attraction forms the interlayer exciton. Similar to the case of intralayer exciton in monolayer TMD [29-31], the spin and valley degree of freedom of interlayer exciton in TMD heterostructure can be addressed using circularly polarized optical excitation [32]. For simplicity, without considering Moire pattern polarization correction [33] (more details in Supplemental Material), here we denote right circularly polarized light ($\sigma_-$) couples to spin up in K valley and left circularly polarized light ($\sigma_+$) couples to spin down in K' valley.

The primary finding is that the energy of the emission from K or K' valley depends on the excitation light polarization. Figure 1(a) shows the interlayer exciton spectrum and peak position under different excitation polarization at $B = 2$ T and temperature 2 K. As shown in Fig. 1(a), with $\sigma_+$ light excitation, light emission with polarization $\sigma_+$ has a lower energy than the light emission with polarization $\sigma_-$ (fitting method shown in Supplemental Material). More interestingly, $\sigma_+$ emission energy becomes higher than $\sigma_-$ emission energy if we use $\sigma_-$ light excitation, as shown in the lower spectrums in Fig. 1(a). We emphasize that we monitor the energy splitting of the $\sigma_+$ and $\sigma_-$ emission but not their intensity [32].

In order to know more about the spectrum splitting, we further measured the cases under a series of different out-of-plane magnetic field. From Fig. 1(b) (red and blue dots), one can see that such spectrum splitting exists in all magnetic field strength (see also Supplemental Material for data from another sample). Even in the absence of magnetic field, such energy splitting exists even though with a smaller value. The splitting here is much larger than the normal valley Zeeman splitting in an external magnetic field as measured in the copolarized cases [12, 13], as denoted with purple dots in Fig. 1(b).

Next, we analyze the mechanism of this optically controlled spectrum splitting and compare with previous pseudo-magnetic field generation schemes. Unlike phenomena observed in TMD monolayer [22, 23], the observed optically induced valley degeneracy breaking cannot be attributed to the valley-selective optical Stark effect. The optical excitation energy (1.707 eV) is too far blue-detuned from the interlayer exciton energy level range (1.3 - 1.4 eV) and its intensity is too small (< 100 kW/cm$^2$). For comparison, the detuning and the intensity of the optical pump used to generate the valley degeneracy breaking of more than 2 meV is around 0.2 eV red-detuned and 1 GW/cm$^2$ respectively [22]. Hence, the contribution of valley-selective optical Stark effect can be neglected.

The emergence of the effective magnetic field using the strong spin-orbit coupling of valence band states in $WSe_2$ is sketched in Fig. 2(a). When we resonantly excite the exciton transition of $WSe_2$ with $\sigma_-$ light, electrons and holes with spin up are generated in $WSe_2$. Due to the energy band alignment, electrons will tunnel to the conduction band of MoSe$_2$ in a short time while spin up holes remain in WSe$_2$, which provides the lowest energy states available for holes. In this way, the electrons and holes are separated ensuring that electron-hole exchange is ineffective in effecting a joint electron-hole valley flip. Strong spin-orbit coupling ensures that holes have a long spin/valley lifetime exceeding microsecond [34]. Due to Coulomb exchange interaction, the presence of optically injected spin-polarized holes imply that in the presence of a large hole population in the K-valley, the net Coulomb repulsion experienced by a K-valley (spin up) hole is reduced as compared to that of a K'-valley hole. As a consequence, the energy of a K-valley interlayer exciton is reduced as compared to that of K'-valley, leading to the observed splitting in the emission spectrum.

The fast change of valley splitting between $B = 0$ T to $B = \pm 2$ T (Fig. 1(b)) is observed for the case of σ– and σ+ excitation but not observed for copolarized case. To explain the rapid increase of the strength of the optically generated field with the applied external B field, we note that the electron-hole exchange interaction induced valley-flip of intra-layer excitons can be suppressed with B fields. Consequently, the increase in hole spin polarization with increasing B field stems from the suppression of hole valley flips taking place before the electron is transferred to MoSe$_2$ [35]. For a magnetic field magnitude bigger than 2 T the expected linear behavior of valley splitting [12, 13] is observed since the interlayer scattering has been largely suppressed. This nonlinear behavior is not observed in copolarized case because the pseudo magnetic field due to two orthogonal optical excitations will have an opposite sign. Hence, the valley splitting is only affected by the external magnetic field for the copolarized case. More evidences that the valley splitting follows valley polarization are shown in Supplemental Material for excitation laser power and wavelength dependent measurement.

We use the rate equation model shown in Fig. 2(b) to obtain an expression for the pseudo magnetic field. The pseudo magnetic field is proportional to the difference between the population of spin up and spin down hole plus the difference between the populations of spin down and spin up electron. Hence it can be written as

$$B \propto \frac{k_e - k_h}{(\gamma + 2k_h)(\gamma + 2k_e)}(\Gamma - \Gamma') \qquad (1)$$

Where $B$ represents the pseudo-magnetic field, $k_e$ ($k_h$) is the depolarization rate of electron (hole), $\Gamma$ ($\Gamma'$) is the electron-hole injection rate in K (K') valley and $\gamma$ is the electron-hole recombination rate.

Based on equation (1), it is possible to tune the valley splitting by changing the proportion of σ+ and σ– part of the optical excitation. In order to prove this, we measure the valley splitting as a function of the polarization state $|\Psi_1\rangle = \frac{1}{\sqrt{2}}(|H\rangle + e^{i\alpha}|V\rangle)$ and $|\Psi_2\rangle = \cos(\frac{\alpha}{2})|H\rangle + ie^{i\phi}\sin(\frac{\alpha}{2})|V\rangle$ with $\phi = -136 \pm 5°$ and various values of $\alpha$. The data is collected for two different values of magnetic field ($B = -2$ T and 2 T). The result is shown in Fig. 3. As shown in this figure, we have successfully controlled the valley degeneracy by excitation polarization manipulation. The data fits well with the pseudo magnetic field model (see Supplemental Material for more detail of the model). Based on this data, by comparing the amplitude of the oscillation with the constant difference between the valley splitting at $B = -2$ T and 2 T, as shown by the difference of two oscillation curves in Fig. 3(c)-(d), we estimated that pseudo magnetic field induced by circularly polarized light is ~ ±31 T.

In summary, we shows that the degeneracy between the valleys can be lifted by using optical excitation even for vanishing external magnetic fields, which is induced by optically induced spin pumping mechanism. Furthermore we show that the valley splitting can be well controlled by manipulating the optical excitation state. The realization of optically induced pseudo-magnetic field will pave the way for optical control of valleytronics devices which is considerably faster than controlling external magnetic field.

We acknowledges the discussion with Marco Battiato, the support from the Singapore National Research Foundation through Singapore NRF fellowship grants (NRF-NRFF2015-03), Competitive Research Programme (CRP Award No. NRF-CRP14-2014-02), Astar QTE and Singapore Ministry of Education (No. MOE2016-T2-2-077, No. MOE2017-T2-1-163) and a start-up grant (No. M4081441) from Nanyang Technological University. Q. H. Xiong acknowledges the support for this work from the Singapore National Research Foundation through an Investigatorship Award (NRF-NRFI2015-03), and Singapore Ministry of Education via two AcRF Tier 2 grants (MOE2015-T2-1-047 and MOE2017-T2-1-040).

## APPENDIX: MATERIALS AND METHODS

The $MoSe_2/WSe_2$ heterostructure is fabricated via mechanical exfoliation and aligned transfer method. The stacking sequence of the sample is checked using second harmonic generation (SHG) based measurement and it is found to be AA stacking. A homemade fiber-based confocal microscope is used for performing the polarization-resolved PL spectroscopy. The polarization state of the excitation and collection is controlled using combination of polarizer, quarter wave plate, and half wave plate. The PL emission is directed by a multi-mode optical fiber into a spectrometer (Andor Shamrock) with a CCD detector for spectroscopic recording. The sample is loaded into a magneto cryostat to control the magnetic field and sample temperature.

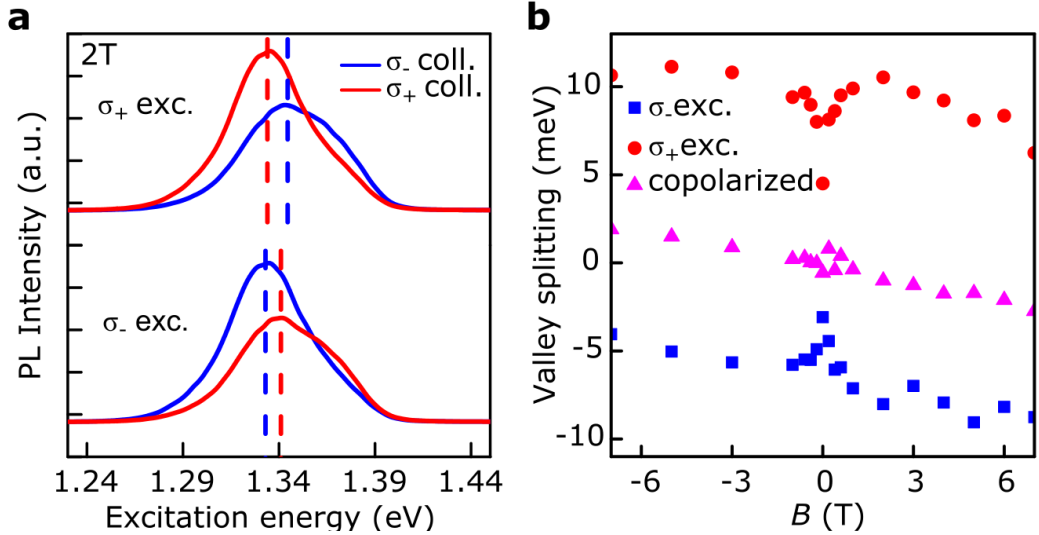

FIG. 1. (a) Interlayer exciton spectrum at 2 K and magnetic field $B = 2$ T for $\sigma_+$ and $\sigma_-$ excitation. The red (blue) solid and dashed lines indicate the spectrum and main peak position (obtained using multiple peak fitting) of $\sigma_+$ ($\sigma_-$) collection respectively. The difference of peak position for the same collection polarization indicates optically induced valley degeneracy breaking. (b) Magnetic field dependence of the valley splitting. The valley splitting is defined as the difference between $\sigma_-$ and $\sigma_+$ collection peak position. Three different cases are shown: $\sigma_-$ excitation, $\sigma_+$ excitation, and copolarized (collection state = excitation state) case. The excitation state dependence of the valley splitting is attributed to optically pseudo magnetic field. The copolarized case shows Zeeman-like linear behavior while the other two show a non-linear trend (a dip) near $B = 0$ T which is attributed to the suppression of intervalley scattering by out-of-plane magnetic field.

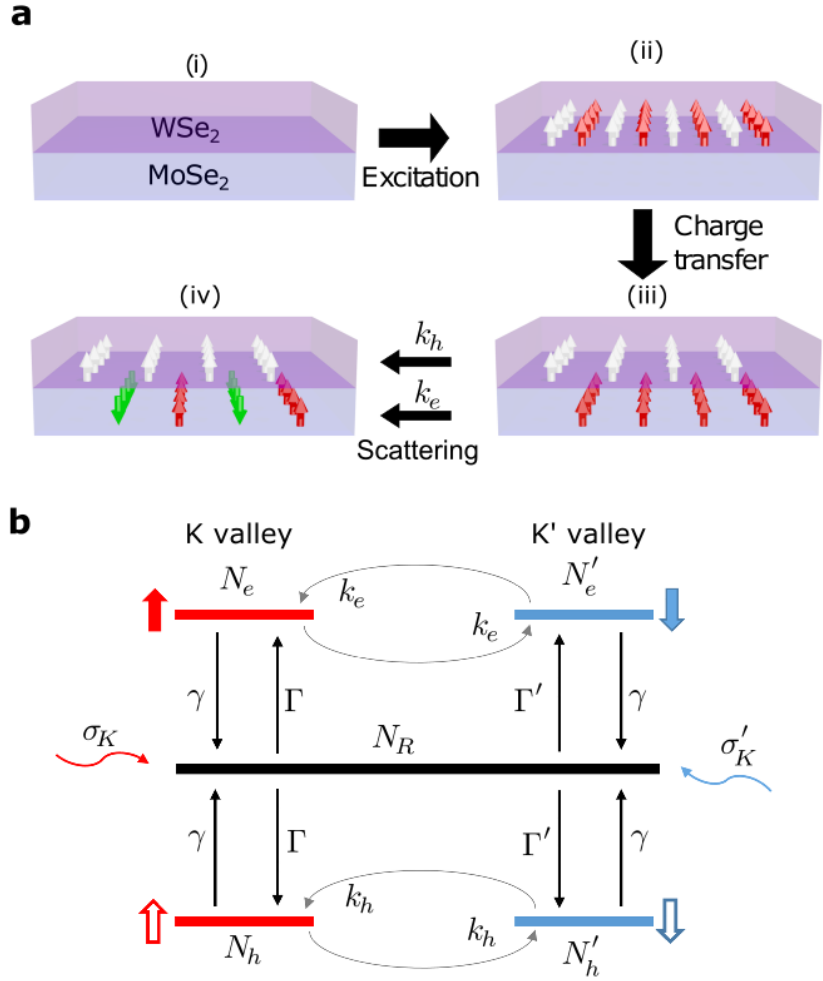

FIG. 2. (a) Schematic of the generation of the non-zero net spin. (i), initial heterostructures. (ii), the heterostructure is excited with circularly polarized laser creating the electron and hole in $WSe_2$ layer. (iii), due to the ultrafast charge (electron) transfer from $WSe_2$ to $MoSe_2$, most of the created conduction band electron migrate to the $MoSe_2$ layer. (iv), both carriers (i.e. hole and electron) undergo depolarization mainly due to the intervalley scattering. Since the electron scattering rate in $MoSe_2$ ($k_e$) is much faster than the hole scattering rate in $WSe_2$ ($k_h$), the system net spin is not zero. This net spin creates a pseudo magnetic field that affects the exciton emission energy. (b) Rate equation model for the process described in **a**. In this model, $N_e$ ($N_e'$) and $N_h$ ($N_h'$) are the electron and hole population in K (K') valley respectively, $N_R$ is non-excited population density, $\Gamma$ ($\Gamma'$) is the electron-hole injection rate in K(K') valley, $\gamma$ is the electron-hole recombination rate, and $\sigma_K$ ($\sigma_K'$) is the optical state that is coupled to K(K') valley.

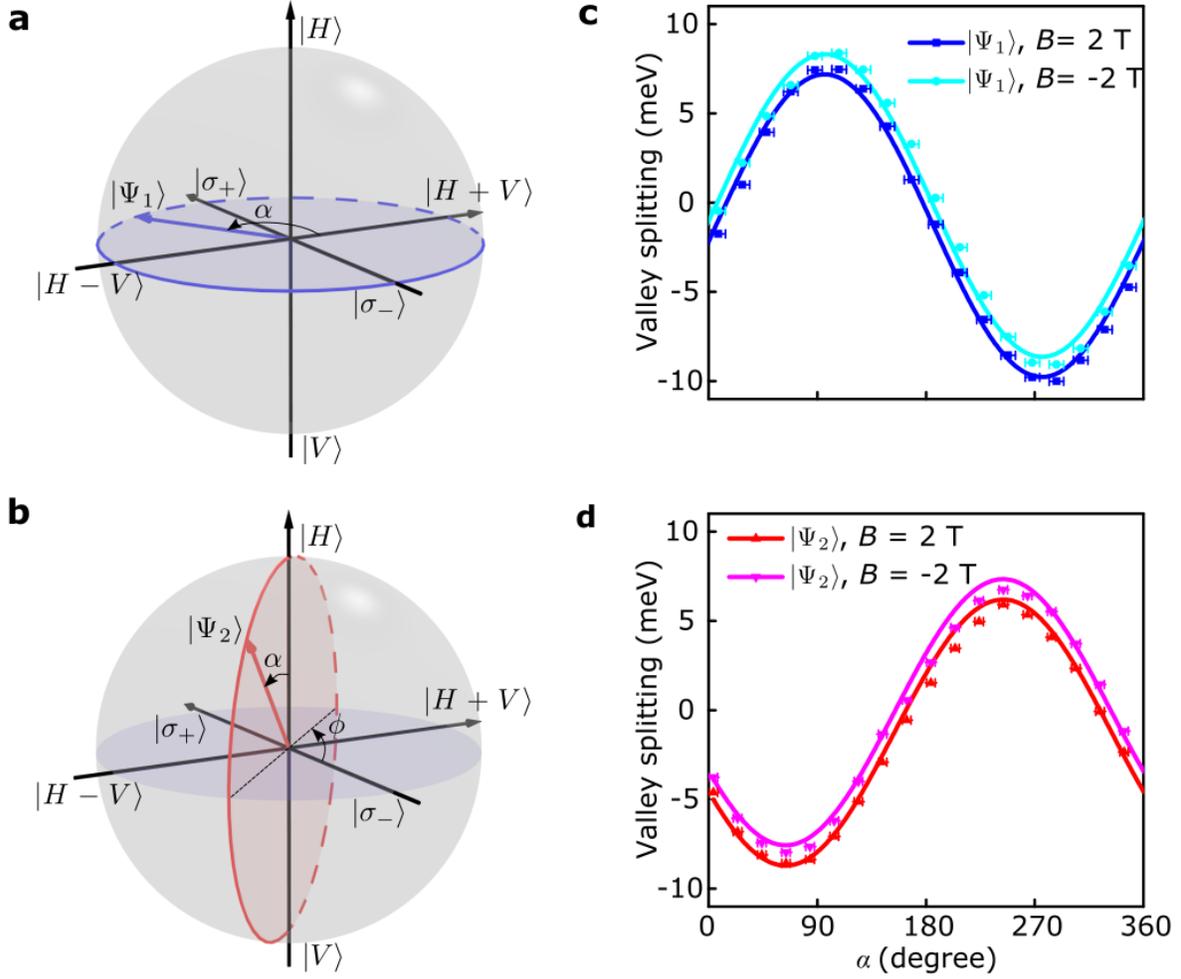

FIG. 3. (a)-(b) Illustration of the trajectory of the optical state on the surface of the Poincaré sphere. The polarization state of the optical excitation is swept through two different trajectories. The state for those trajectories can be expressed as $|\Psi_1\rangle = \frac{1}{\sqrt{2}}(|H\rangle + e^{i\alpha}|V\rangle)$ and $|\Psi_2\rangle = \cos(\frac{\alpha}{2})|H\rangle + ie^{i\phi}\sin(\frac{\alpha}{2})|V\rangle$ with $\phi = -136 \pm 5°$ for trajectory in (a) and (b) respectively. (c)-(d) Valley splitting as a function of excitation polarization state. Here the dots represent the experimental data while the lines are the theoretical fitting result. The result in (c) corresponds to the trajectory in (a), while (d) corresponds to (b).

# Supplemental Material: Optical spin pumping induced pseudo-magnetic field in two dimensional heterostructures

Chongyun Jiang[1†], Abdullah Rasmita[1†], Weigao Xu[1], Atac Imamoğlu[2], Qihua Xiong[1,3,4*], Wei-bo Gao[1,4,5*]

[1]Division of Physics and Applied Physics, School of Physical and Mathematical Sciences, Nanyang Technological University, Singapore 637371, Singapore.

[2]Institute of Quantum Electronics, ETH Zürich, CH-8093 Zürich, Switzerland

[3]NOVITAS, Nanoelectronics Center of Excellence, School of Electrical and Electronic Engineering, Nanyang Technological University, Singapore 637371, Singapore.

[4]MajuLab, CNRS-Université de Nice-NUS-NTU International Joint Research Unit UMI 3654, Singapore.

[5]The Photonics Institute and Centre for Disruptive Photonic Technologies, Nanyang Technological University, Singapore 637371, Singapore.

*Corresponding author. Email: qihua@ntu.edu.sg; wbgao@ntu.edu.sg

†These authors contributed equally to this work

## I. Valley splitting dependence on excitation state

In 2D transition metal dichalcogenide (TMD) heterostructure, K and K' valley is not necessarily coupled to $\sigma_-$ or $\sigma_+$ excitation due to the lattice mismatch between the two layer [1, 2]. Due to the Moiré pattern created, the optical selection rule and the oscillator strength becomes position dependent with periodicity around few nm [1]. Since our excitation beam has diameter in μm scale, the optical polarization state that has largest coupling strength to the K valley will in general have elliptical polarization. We denote this optical polarization state as $\sigma_K$. Due to the time reversal symmetry, the optical polarization state that has largest coupling strength to the K' valley (denoted as $\sigma_K'$) will be orthogonal to $\sigma_K$.

We define the valley splitting as the energy difference between the emission from K valley (spin up electron, $E_K$) and K' valley (spin down electron, $E_K'$). Since $E_K$ ($E_K'$) will decreases (increases) with increasing pseudo magnetic field, the valley splitting has opposite sign compared to pseudo magnetic field. For $|\sigma_K\rangle = \begin{bmatrix} \cos(\frac{\theta}{2}) \\ e^{i\varphi}\sin(\frac{\theta}{2}) \end{bmatrix}$, the valley splitting $VS$ caused by the optical excitation with polarization state $|\Psi\rangle = \begin{bmatrix} \cos(\frac{\theta'}{2}) \\ e^{i\alpha}\sin(\frac{\theta'}{2}) \end{bmatrix}$ can be expressed as

$$VS(|\Psi\rangle) \propto \left|\langle\sigma'_K|\Psi\rangle\right|^2 - \left|\langle\sigma_K|\Psi\rangle\right|^2$$
$$\propto -\cos(\theta)\cos(\theta') - \sin(\theta)\sin(\theta')\cos(\alpha - \varphi)$$

The $|\sigma_K\rangle$ state can be obtained by fitting the valley splitting expression above to the valley splitting data for the two circular trajectories in Poincaré sphere (see main text Fig. 3) with a constant term is added to represents the valley splitting due to the external magnetic field. Based on this fitting result, we obtained $\theta = 68 \pm 6°$ and $\varphi = -84 \pm 6°$ where the uncertainty is derived from the uncertainty in the orientation of the optical components.

We note here that the optical selection rule has position dependence owing to the Moire pattern and the inhomogeneity of the sample. In order to show this, the optical control of valley splitting experiment is conducted at different position on the same sample. The result is shown in Fig. S5(a)-(f). As can be seen from these figures, the result is different compared to the one in Fig. 3 in the main text. This is especially clear from Fig. S5(d) where it is shown that the valley splitting does not depend on the excitation polarization when the excitation state is on the plane normal to σ− and σ+ (see Fig. S5(b) for an illustration of the trajectory). This shows that, at this different position, the K(K') valley is coupled to σ−(σ+) excitation.

## II. Multiple peaks fitting of interlayer exciton emission

We fit the interlayer exciton spectrum with multiple peaks Gaussian fitting. It is found that the spectrum consists of two prominent peaks (Fig. S1). The origin of these two peaks can be attributed to the different interlayer exciton energy for different stacking configuration which should exist because of the lattice mismatch between the $MoSe_2$ and $WSe_2$ [1]. In our study we only put concern on the peak with the largest intensity (peak 1, energy ~ 1.34 eV). Multiple peak fitting has been used in magnetic field dependence in Fig. 1 and Fig. S2. In most of our data, the second peak is small enough to be ignored. Hence, in order to avoid error due to overfitting, single peak fitting is used to obtain the results shown in Fig. 3 and Fig. S3-5.

## III. Additional data from different sample

In order to test the repeatability of the optically induced pseudo magnetic field, we performed the magneto-optic experiment on another $MoSe_2/WSe_2$ sample. The magnetic and excitation polarization dependence of the valley splitting for this sample is shown in Fig. S2. As can be seen from Fig. S2, the optically induced pseudo magnetic field is also observed for this sample.

## IV. Power, excitation wavelength, and temperature dependence

The temperature dependence of the valley splitting at $B = 2$ T is shown in Fig. S3. The effect of the excitation polarization to the valley splitting can be observed at temperature as high as 90 K. Its effect decreases with increasing temperature. This can be understood as the valley depolarization rate is higher at higher temperature [3] and, hence, the optically induced pseudo magnetic field will also decrease with increasing temperature.

The power and excitation wavelength dependence of the valley splitting and degree of polarization at $B = 2$ T are shown in Fig. S4(a)-(d). Here, the degree of polarization is defined as $\frac{I(\sigma_-) - I(\sigma_+)}{I(\sigma_-) + I(\sigma_+)}$ where $I(\psi)$ is the intensity of collected light projected to state $\psi$. Comparing Fig. S4(a)-(b) and S4(c)-(d), it can be seen that there is a close relationship between the valley polarization and valley splitting. This is expected since the pseudo magnetic field depends on the net spin polarization which in turn depends on the valley polarization.

Unlike in the case of valley splitting due to the optical Stark effect or Bloch-Siegert shift, the observed valley splitting does not increase with increasing excitation power (Fig. S4(a)). Other than the apparent decreasing trend that can be attributed to optical excitation-induced heating, the valley splitting tends to be a constant of power. This can be understood as the pseudo magnetic field only depends on the population density difference between spin up and spin down hole. The hole population density tends to reach a saturation point. Hence, the population density difference between spin up and spin down hole only depends on the ratio of between electron-hole injection rate in K(K') valley and this ratio does not have strong dependence on power.

The effect of optically-induced valley splitting is maximum near 726 nm (Fig. S4(c)) which corresponds to the $WSe_2$ charged exciton energy level. At 532 nm, there is virtually no difference between different excitation. This can be understood as the pseudo magnetic field depends on the polarized hole population generation in $WSe_2$ layer which efficiency reduced as the excitation energy differ from the $WSe_2$ exciton energy level.

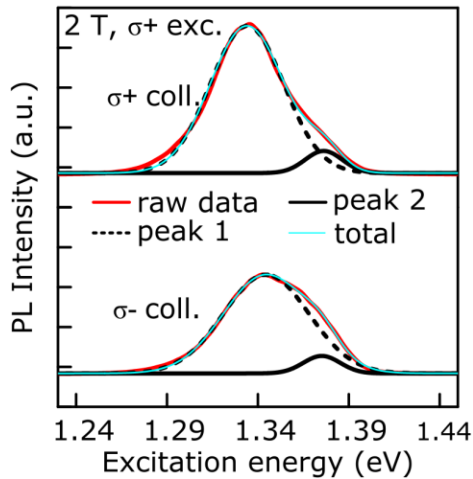

FIG. S1. Multiple peaks fitting of interlayer exciton at $B = 2$ T and temperature 2 K. The two-peaks Gaussian fitting result (light-blue solid lines) shows a good agreement with the raw data (solid red lines). The parts of the spectrum corresponded to the first and the second peak are shown as dotted and solid black lines respectively.

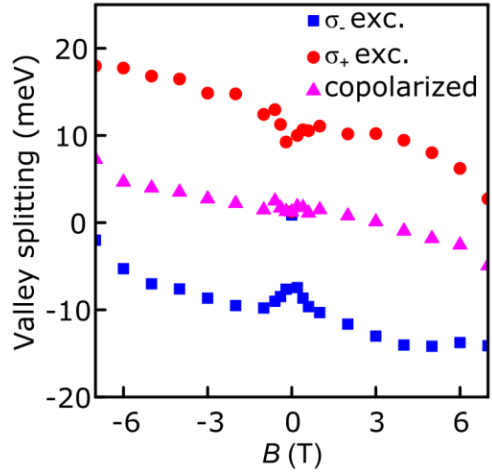

FIG. S2. Excitation polarization and magnetic field dependence of the interlayer exciton valley splitting for second MoSe$_2$/WSe$_2$ sample. A similar dependence is observed for the second sample which shows the repeatability of the observed optically induced pseudo magnetic field.

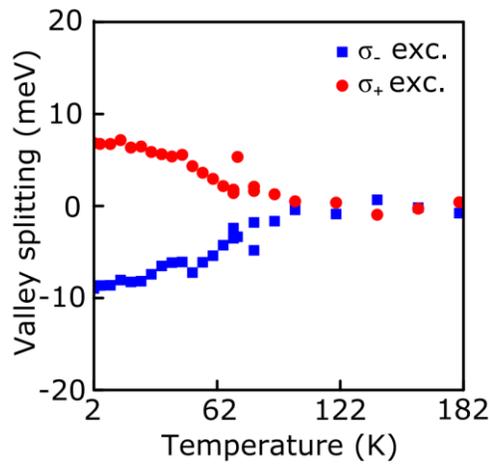

FIG. S3. Temperature dependence. Temperature dependence of valley splitting under $B = 2$ T with excitation power 510 µW and excitation wavelength 726 nm.

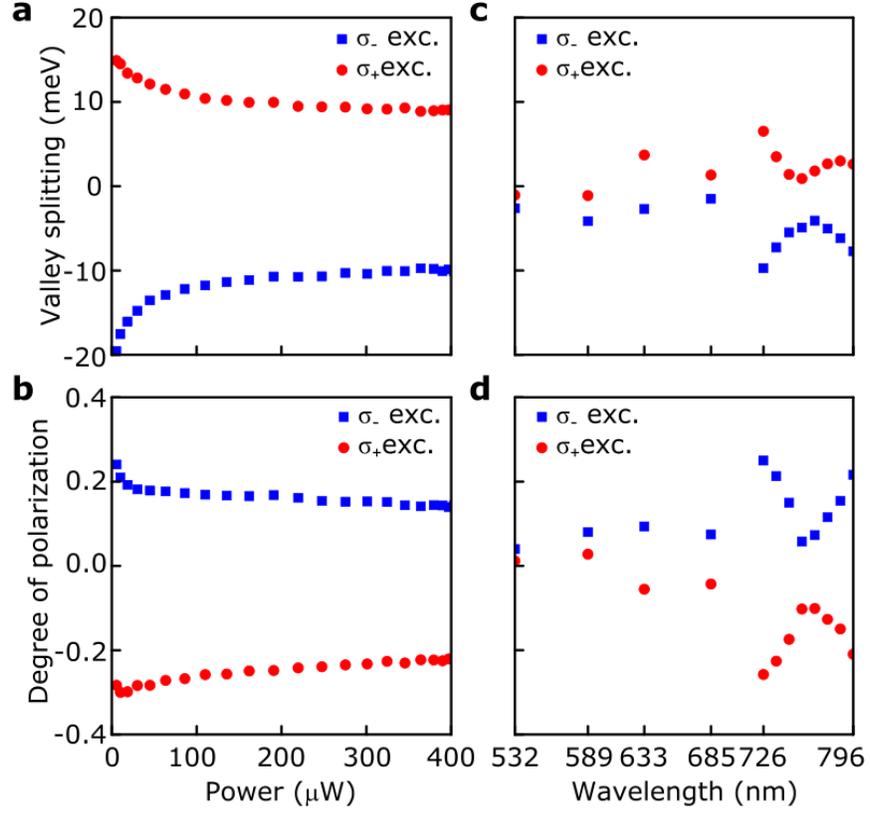

FIG. S4. Power and excitation wavelength dependence. (a)-(b) Power dependence of valley splitting and degree of polarization (DoP) under $B = 2$ T with excitation wavelength 726 nm and temperature 2 K. The degree of polarization is defined as $\frac{I(\sigma_-) - I(\sigma_+)}{I(\sigma_-) + I(\sigma_+)}$ where $I(\psi)$ is the intensity of collected light projected to state $\psi$. (c)-(d) Excitation wavelength dependence of the valley splitting and DoP at $B = 2$ T and temperature 2 K. The powers at each wavelength are 260 μW for 532 nm, 140 μW for 589 nm, 455 μW for 633 nm, 385 μW for 685 nm, and 460 μW for 726-796 nm.

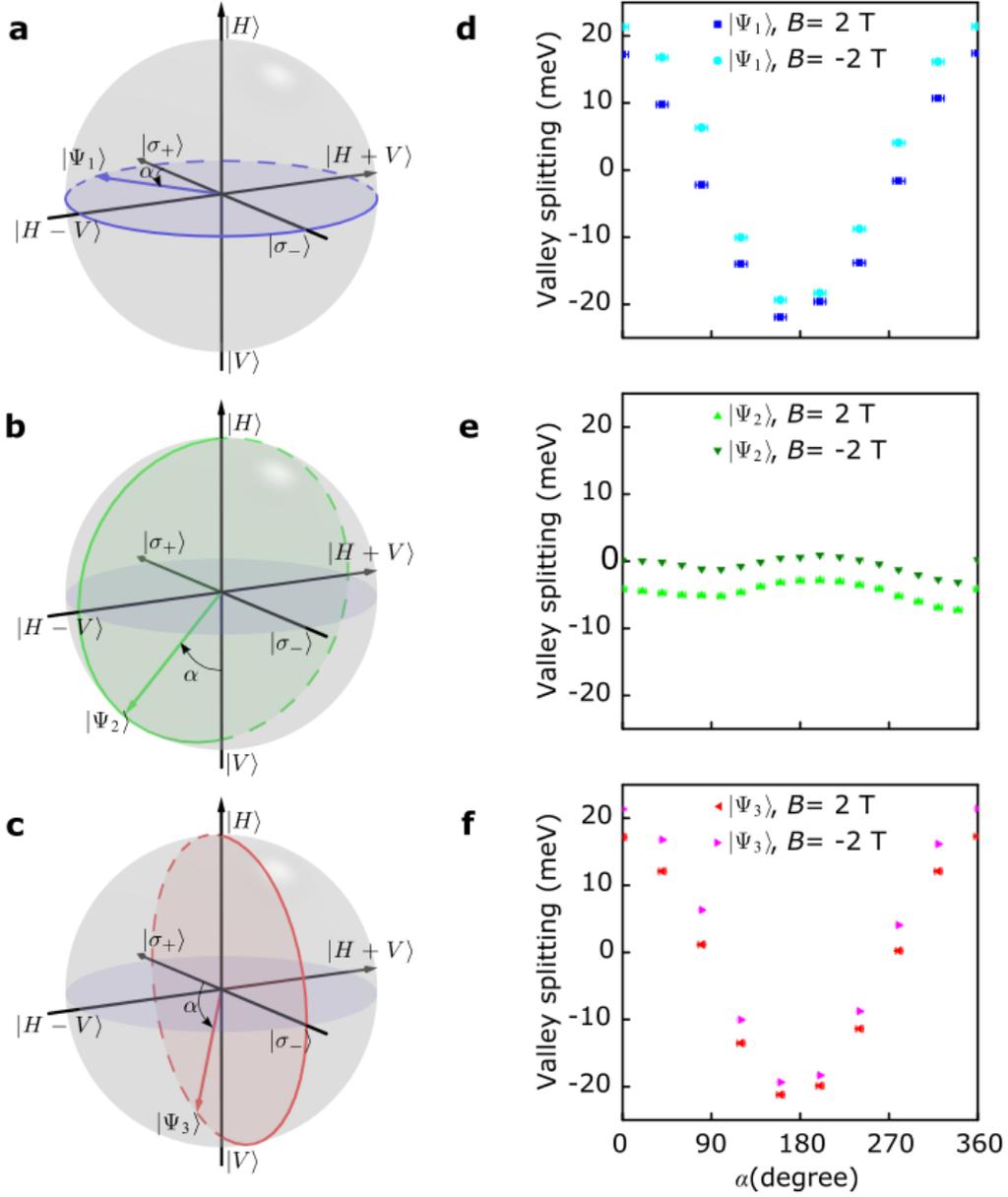

FIG. S5. All optical control of valley splitting at a different position. (a)-(c) Illustration of the trajectory of the optical state on the surface of the Poincaré sphere. The states for those trajectories are $|\Psi_1\rangle = \frac{1}{\sqrt{2}}(|H\rangle + ie^{i\alpha}|V\rangle)$, $|\Psi_2\rangle = \sin(\frac{\alpha}{2})|H\rangle - \cos(\frac{\alpha}{2})|V\rangle$, and $|\Psi_3\rangle = \cos(\frac{\alpha}{2} + \frac{\pi}{4})|H\rangle + i\sin(\frac{\alpha}{2} + \frac{\pi}{4})|V\rangle$ for trajectory in (a), (b) and (c) respectively. (d)-(f) Valley splitting as a function of excitation polarization state. The result in (d), (e), and (f) corresponds to the trajectory in (a), (b) and (c) respectively.